\documentclass[floatfix,prl,superscriptaddress,twocolumn]{revtex4}
\usepackage{graphicx}
\usepackage{graphics}
\usepackage{footmisc}
\usepackage{latexsym}
\usepackage{amsmath, amsthm, amssymb, wasysym}
\usepackage{xcolor}
\usepackage{epsfig}

\begin{document}
\title{General approach to stochastic resetting}

\author{R. K. Singh}
\email{rksinghmp@gmail.com } \affiliation{Department of Physics, Bar-Ilan University, Ramat-Gan 5290002, Israel}

\author{K. G\'orska}
\email{katarzyna.gorska@ifj.edu.pl} \affiliation{Institute of Nuclear Physics, Polish Academy of Sciences, PL-31342 Krak\'ow, Poland
}

\author{T. Sandev}
\email{trifce.sandev@manu.edu.mk} \affiliation{Research Center for Computer Science and Information Technologies, Macedonian Academy of Sciences and Arts, Bul. Krste Misirkov 2, 1000 Skopje, Macedonia}
\affiliation{Institute of Physics \& Astronomy, University of Potsdam, D-14776 Potsdam-Golm, Germany}
\affiliation{Institute of Physics, Faculty of Natural Sciences and Mathematics, Ss.~Cyril and Methodius University, Arhimedova 3, 1000 Skopje, Macedonia}

\date{\today}

\begin{abstract}
We address the effect of stochastic resetting on diffusion and subdiffusion process. For diffusion we find that MSD relaxes to a constant only when the distribution of reset times possess finite mean and variance. In this case, the leading order
contribution to the PDF of a Gaussian propagator under resetting exhibits a cusp independent of the specific details of the reset time
distribution.
For subdiffusion we derive the PDF in Laplace space for arbitrary resetting protocol. Resetting at constant rate
allows evaluation of the PDF in terms of $H$-function. We analyze the steady state and derive the rate function governing the relaxation behavior. For a subdiffusive process the steady state could exist even if the distribution of reset times possesses only finite mean.
\end{abstract}

\maketitle

\newcommand{\tl}{\tilde}
\newcommand{\avg}{\langle \tau \rangle}
\newcommand{\var}{\langle \tau^2 \rangle}
\newcommand{\msdlr}{\langle \tilde{x}^2_r(s) \rangle}
\newcommand{\msdlt}{\langle {x}^2_r(t) \rangle}
\newcommand{\msdt}{\langle x^2(t) \rangle}
\newcommand{\fr}{\frac}

\textit{Introduction}: 
Stochastic resetting is a ubiquitous phenomena manifesting itself in
random search processes. For example, whenever an animal goes out in search
of food, it is generally not successful in the first attempt, failing which
it retrieves back home. This is not just typical of animal foraging, but generic
to any search process wherein it culminates following the find, but
is interrupted every now and then via restarts \cite{jparev}. And in some cases,
like in computer science, random restarts are known to optimize search algorithms
\cite{luby, montanari}. A paradigmatic example of search process in statistical
physics is Brownian motion in presence of an absorbing wall \cite{redner}
and resetting the motion at constant rates renders mean time finite \cite{evans}.
In Ref.~\cite{reuveni} Reuveni shows that for an optimally restarted process
at fixed rate the signal to noise ratio is unity, a result which is further extended
by Pal and Reuveni \cite{pal1} to time-dependent restart rates \cite{pal2}.
These results have also been extended
to address random search via renewal by Chechkin and Sokolov \cite{chechkin}
who show that resetting is always useful if probability of finding a target
in absence of restarts decays slower than exponential.

In addition to having a significant effect on first passage properties, stochastic
resetting also affects transport behavior. This is visibly seen, for example, in case
of diffusive processes. Resetting a Brownian particle at constant rate results in a
nonequilibrium steady state (NESS) at long times \cite{schehr}. Such resetting-induced
NESS have also been observed for coagulation-diffusion processes \cite{park}, height
distribution of fluctuating interfaces \cite{gupta}, etc. Existence of a NESS
is not independent of the behavior of the mean square displacement (MSD) of a stochastic
process under reset. Motivated by this observation, Bodrova and co-workers study
MSD under renewal and nonrenewal resetting for scaled Brownian motion and show
that the properties of probability distribution for renewal and nonrenewal case
are vastly differently \cite{bodrova1, bodrova2}. Transport properties are also
addressed by Mas\'{o}-Puigdellosas and co-workers for a general random walk \cite{campos}.
However, their analysis is limited to the case when the distribution of reset times
has a specific form, like exponential or power law, or Mittag-Leffler which connects
the two \cite{klafter}. In this work we generalize to the case when the distribution
of reset times either possesses both mean and variance, or one of them, or none, without
making any other assumptions about the nature of the underlying distribution.

For the case of linear diffusion we show that a 
condition for the existence of a steady state under stochastic reset is that the distribution of reset times possesses both mean and variance. In the case of sub-diffusion the steady states under stochastic reset is observed when only the mean value exist.

The paper is organized as follows: first we discuss the effect
of resetting on diffusion. Next
we study how resetting modifies the probability distribution of a Gaussian propagator,
followed by a generalization to subdiffusive transport via subordination. For resetting
at constant rate we derive the explicit form of probability distribution and address
the relaxation behavior towards steady state generalizing earlier result on combs
\cite{rks}.

\textit{Stochastic resetting}:
Let the motion of a particle starting at $x = 0$ at $t = 0$ be described
by the propagator $p(x,t)$, which is renewed at random times following the
distribution $r(t)$. Then the dynamics of the particle under stochastic
resetting is described by the renewal equation \cite{campos}
\begin{align}
\label{renewal}
p_r(x,t) = R(t)p(x,t) + \int^t_0 dt' r(t')p_r(x,t-t'),
\end{align}
where $R(t) = \int^\infty_t dt' r(t')$ is the probability that no renewal
has taken place upto time $t$. By (\ref{renewal}) we have transport behavior
under stochastic reset,
\begin{align}
\msdlr = \fr{\mathcal{L}\left\{R(t)\msdt\right\}}{1-\tl{r}(s)},
\end{align}
where $\mathcal{L}\left\{f(t)\right\} = \tl{f}(s)$ denotes Laplace transform.

\textit{Diffusive transport}:
Let us focus now on the effect of stochastic resetting on diffusive transport,
that is $\msdt = 2t$. This form of transport is generic to many stochastic processes \cite{book_diffusion}, and hence we focus on this case.
For the resetting protocol, we do not make any assumptions on the nature
of distribution of reset times $r(t)$ except its behavior at long times. In order to
address this, we choose, for example see \cite{godreche},
\begin{align}\label{reset_distribution}
\tl{r}(s) \approx \begin{cases}
1 - \avg s + \fr{1}{2}\var s^2, ~\text{mean and variance exist},\\
1 - \avg s + a s^\theta, ~\text{mean exists, $1 < \theta <2$},\\
1 - as^\theta, ~~\text{mean does not exist, $0 < \theta < 1$},
\end{cases}
\end{align}
where $a=|\Gamma(1-\theta)|\tau_{0}^{\theta}$, and $\tau_0$ is the microscopic time scale. 
For the case when the distribution of reset times has both finite mean and variance (narrow distribution), that is $\theta > 2$ we have, $\tl{R}(s) = [1-\tl{r}(s)]/s \approx \avg - \var s/2$, resulting in $\msdlr \stackrel{s \rightarrow 0}{\approx} \var/(\avg s)$. This implies that at long times the MSD saturates, i.e.,
\begin{align}
\msdlt = \var/\avg.
\end{align}
For the case $1 < \theta < 2$ (broad distribution) the MSD becomes $ \msdlr \approx 2a(\theta-1)s^{\theta-2}/(\avg s - as^{\theta}) \stackrel{s \rightarrow 0}{\approx} \fr{2a(\theta-1)}{\avg}\fr{1}{s^{3-\theta}}$. Hence the long time behavior of MSD is subdiffusive
\begin{align}
\msdlt = \fr{2a(\theta-1)}{\avg\Gamma(3-\theta)}t^{2-\theta}.
\end{align}
For $0<\theta<1$ (broad distribution) we find
$\msdlr \approx -\fr{2a(\theta-1)s^{\theta-2}}{a s^{\theta}} = \fr{2(1-\theta)}{s^2}~\Rightarrow\msdlt = 2(1-\theta)t$. This implies that stochastic resetting does not change the nature of transport
if the distribution of resetting times does not posses a mean, in that the motion remains diffusive even after resetting. However, the diffusion coefficient for motion under stochastic reset is less than unity, and that is the signature of resetting.
For the marginal cases $\theta = 1$ and $\theta = 2$, the large time behavior of reset
time distribution in Laplace domain is \cite{badii}
\begin{align}
\tl{r}(s) \stackrel{s \rightarrow 0}\approx \begin{cases}
1 + s\ln s, & \theta = 1,\\
1 - \avg s - s^2 \ln s, & \theta = 2.
\end{cases}
\end{align}
Thus, by an application of the Tauberian theorems \footnote{If a function $r(t)$, $t\geq 0$, has the Laplace transform $\tl{r}(s)$ whose asymptotics behaves as follows
$\tl{r}(s)\sim s^{-\rho}L\left(\frac{1}{s}\right), \quad
s\rightarrow0, \quad \rho>0,$ then
$r(t)=\mathcal{L}^{-1}\left\{\tl{r}(s)\right\}\sim
\frac{1}{\Gamma(\rho)}t^{\rho-1}L(t)$, for $t\rightarrow\infty$. Here $L(t)$ is a slowly varying function at infinity, i.e., $\lim_{t\rightarrow\infty}L(at)/L(t)=1,$ for any $a>0$ \cite{feller}.} we have
$\msdlt = 2t/\ln t$ for $\theta = 1$ and $\msdlt = 2\ln t/\avg$ for $\theta=2$.

In summary, if MSD at long times is $\msdt \sim 2t$,
then stochastic resetting modifies transport behavior as follows:
\begin{align}
\label{msd}
\msdlt =\begin{cases}
2(1-\theta)t, & 0< \theta < 1,\\
2t/\ln t, & \theta = 1,\\
\fr{2a(\theta - 1)}{\avg \Gamma(3-\theta)} t^{2-\theta}, & 1< \theta < 2,\\
2 \ln t/\avg, & \theta = 2,\\
\var/\avg, & \theta > 2.
\end{cases}
\end{align}
From (\ref{msd}) it also becomes clear that at long times the MSD
under resetting $\msdlt$ saturates to a constant value only when the distribution
of reset times possess both finite mean and variance. This is a necessary condition
for the existence of a steady state under stochastic resetting.


\textit{Effect of resetting on PDF}:
Let us assume that the particle evolves in absence of resetting following a Gaussian
probability density function (PDF), that is
$p(x,t) = \fr{1}{\sqrt{4\pi t}}\exp\left(-\frac{x^2}{4t}\right) \iff \hat{p}(k,t) = \exp\left(-k^2 t\right)$,
where $\hat{p}$ denotes Fourier transform. Using this in (\ref{renewal}) we have in Fourier-Laplace space
\begin{align}
\hat{\tl{p}}_r(k,s) = \fr{\mathcal{L}\left\{R(t)\exp\left(-k^2 t\right)\right\}}{1-\tl{r}(s)} =
\fr{\tl{R}(s+k^2)}{1-\tl{r}(s)}.
\end{align}
Let us now assume that $s$ and $k$ are small in what follows, so that
we can use the small-$s$ expansion of the distribution of restart times
stated in Eq.~(\ref{reset_distribution}). This allows us to address the
effect of stochastic resetting on a Gaussian propagator to leading order.
For narrow distribution 
of reset times when both mean and variance exist, that is, $\theta > 2$, the PDF becomes
\begin{align*}
    \hat{\tl{p}}_r(k,s) \approx \fr{\avg - \fr{1}{2}\var (s+k^2)}{ \avg s - \fr{1}{2}\var s^2} \stackrel{s \rightarrow 0}{\approx} \frac{1}{s}\left(1 - \frac{\var}{2\avg}k^2\right),
\end{align*}
which for small $k$ can be treated as the first terms of $[1 + \var k^2/(2\avg)]^{-1}$. Thus, we have
\begin{align*}
 \hat{\tl{p}}_r(k,s) \stackrel{s \rightarrow 0}{\approx} \fr{1}{s}\fr{1}{1 + \fr{\var}{2\avg}k^2}.
\end{align*}
Inverting the Fourier-Laplace transform we have
\begin{align}
\label{pst}
p_{r}(x,t) 
\approx \fr{1}{2}\sqrt{2\avg/\var}\exp\left(-|x|\sqrt{2\avg/\var}\right).
\end{align}
This implies that to a leading order the PDF assumes a
cusp at the reset location, independent of the specific details of the
distribution of reset times as long as it possesses a finite mean and
variance. It is to be further noted that this leading order term is
time-independent, and represents the steady state for a Gaussian propagator
reset at constant rates.



For the case when the distribution of reset times possesses a finite mean, that is $\theta\in(1, 2)$,
we have $\tl{r}(s) \stackrel{s \rightarrow 0}{\approx} 1 - \avg s + a s^\theta$,
resulting in
$
\hat{\tl{p}}_r(k,s) \approx \fr{\avg - a(s+k^2)^{\theta - 1}}{\avg s - a s^\theta}
\approx \fr{1}{s}\fr{1}{1 + \fr{a}{\avg}\left(s+k^2\right)^{\theta - 1}}$. This implies $\tl{p}_r(k,t) \approx \int^t_0 dt' f(k,t')$ where $\tl{f}(k,s) = \fr{1}{1+\fr{a}{\avg}(s+k^2)^{\theta - 1}}$. By inverse Laplace transform we find 
\begin{align}
    f(k,t) 
    =\fr{\avg}{a}e^{-k^2t}\,t^{\theta-2}E_{\theta-1,\theta-1}\left(-\fr{\avg}{a}t^{\theta-1}\right),\nonumber
\end{align}
where $E_{\alpha,\beta}(z)=\sum_{n=0}^{\infty}\frac{z^n}{\Gamma(\alpha n+\beta)}$ is the two parameter Mittag-Leffler function \cite{ml_paper} (it is a special case of the three parameter Mittag-Leffler function $E_{\alpha, \beta}^{\gamma}(\sigma)$ for $\gamma=1$, see below). By asymptotic expansion of the Mittag-Leffler function in $f(k,t)$ in the long time limit, and by applying $\msdlt=\left(-\frac{\partial^2}{\partial k^2}\hat{p}_{r}(k,t)\right)_{k=0}$, we again obtain the MSD (\ref{msd}) for $1<\theta<2$.

For the special case of $\theta = 3/2$, the inverse Laplace transform evaluates to
\cite{badii}
\begin{align}
    f(k,t) 
    = be^{-k^2 t}\left[\fr{1}{\sqrt{\pi t}}-be^{b^2t}\text{erfc}\left(b\sqrt{t}\right)\right],\nonumber
\end{align}
where $b = \avg/a$ and $\text{erfc}(z)=\frac{2}{\sqrt{\pi}}\int_{z}^{\infty}e^{-y^2}dy$ is the  the complementary error function. As a result, the PDF becomes
\begin{align}
p_r(x,t) & 
\approx \fr{b}{2\pi}\Gamma\Big(0,\fr{x^2}{4t}\Big) - b^2\int^t_0 dt' ~p(x,t')e^{b^2 t'}\text{erfc}(b\sqrt{t'}),
\end{align}
where $\Gamma(\lambda,z)=\int_{z}^{\infty}y^{\lambda-1}e^{-y}dy$ represents the upper incomplete gamma function.

Finally when the distribution of reset times does not even possess the mean, we have
$\tl{r}(s) \stackrel{s \rightarrow 0}{\approx} 1 - a s^\theta$ with $0 < \theta < 1$. This
results in
$
\hat{\tl{p}}_r(k,s) \approx \fr{1}{s^\theta}\fr{1}{\left(s+k^2\right)^{1-\theta}}.
$
The inverse Laplace transform yields \footnote{\label{refnote}The Laplace transform of the three parameter Mittag-Leffler function reads $$\mathcal{L}\left\{t^{\beta-1}\,E_{\alpha, \beta}^{\gamma}(-\lambda t^\alpha)\right\}(s) = \frac{s^{\alpha\gamma-\beta}}{\left(s^\alpha+\lambda\right)^{\gamma}}, \quad \Re(s)>|\lambda|^{1/\alpha}.$$.}
$
    \hat{p}_r(k,t)\approx E_{1,1}^{1-\theta}\left(-k^2t\right)=\phi\left(1-\theta,1,-k^2t\right),
$
where $E_{\alpha,\beta}^{\gamma}(z)=\sum_{k=0}^{\infty}\frac{(\gamma)_k}{\Gamma(\alpha k+\beta)}\frac{z^k}{k!}$ is the three parameter Mittag-Leffler function \cite{ml_paper}, $(\gamma)_{k}$ is the Pochhammer symbol, and $\phi(a,b,z)$ is the Kummer’s confluent hypergeometric function. From here, by inverse Fourier transform, we find
\begin{align}
    p_r(x,t)&\approx
\frac{1}{\Gamma(1-\theta)|x|}H_{2,2}^{2,0}\left[\frac{x^2}{t}\left|\begin{array}{c
l}
    (1,1), (1,1)\\
    (1,2), (1-\theta,1)
  \end{array}\right.\right],
\end{align}
where $H_{p,q}^{m,n}(z)$ is the Fox $H$-function \cite{saxena_book}. We note here that from this PDF, by using the Mellin transform of the Fox $H$-function, for the MSD we again obtain the result given in~(\ref{msd}) for $0<\theta<1$.

For the special case $\theta = 1/2$ it reduces to \cite{badii}
$\hat{p}_r(k,t) =\mathcal{L}^{-1}\left[\fr{1}{\sqrt{s}}\fr{1}{\sqrt{s+k^2}}\right]= e^{-k^2t/2}I_0\left(k^2t/2\right)$, i.e., $\hat{p}_r(k,t)= e^{-k^2t/2}\sum^\infty_{n=0} \fr{1}{(n!)^2}\left(\fr{k^4t^2}{16}\right)^n,
$
which implies
\begin{align}
p_r(x,t) 
\approx \left[1+\sum^\infty_{n=1}\fr{1}{(n!)^2}\left(\fr{t^2}{16}\fr{d^4}{dx^4}\right)^n\right]\fr{e^{-\fr{x^2}{2t}}}{\sqrt{2\pi t}}.
\end{align}
The effect of resetting (even when a mean does not exist) is to reduce the
spread of Gaussian from $2t$ to $t$ ($\theta = 1/2$) in the simplest approximation.

It is to be noted that while our analysis has been limited to the leading order term in the Fourier-Laplace expansion of $\hat{\tl{p}}_r(k,s)$, we can nevertheless say that
$\lim_{t\rightarrow \infty}p_r(x,t)$ converges to a steady state only when $\theta>2$.
This is not directly inconsistent with the analysis of Ref.~\cite{nagar} wherein the authors
find a nontrivial value for $\lim_{t\rightarrow \infty}p_r(x,t)$ for $\theta \in (1,2)$
and refer to it as a steady state. In such a state, however, the particle spends most of
its time far from the reset location characterized by a divergent MSD. This is straightforwardly
seen from Eq.~(\ref{msd}) for $\theta \in (1,2)$.

\textit{Anomalous diffusion with resetting}:
Next we consider a generalized anomalous diffusive process \cite{fcaa2015} 
\begin{align}\label{diff eq Ps sub}
    \int_{0}^{t}\gamma(t-t')\frac{\partial}{\partial t'}P_{\text{s}}(x,t')dt'=L_{\text{FP}}P_{\text{s}}(x,t)
\end{align}
with the memory kernel $\gamma(t)$. It is subordinated to normal diffusion process with the subordination function, which in the Laplace space reads
\begin{align}\label{sub function}
    \tl{h}(u,s)=\tl{\gamma}(s)e^{-us\tl{\gamma}(s)}.
\end{align}
The memory kernel is such that in the long time limit it goes to zero, $\lim_{t\rightarrow\infty}\gamma(t)=0$, i.e., $\lim_{s\rightarrow0}s\tl{\gamma}(s)=0$. It is known that the PDF of this generalized diffusion equation can be obtained as a subordination integral \cite{metzler_prl}
\begin{align}\label{sub integral}
    p(x,t)=\int_{0}^{\infty}p_0(x,u)h(u,t)\,du,
\end{align}
where $p_0(x,u)$ is Gaussian PDF.
Therefore, in presence of resetting, the PDF becomes
\begin{align}
\label{prks}
\tl{p}_r(x,s) = \fr{\mathcal{L}\left\{R(t)p(x,t)\right\}}{1-\tl{r}(s)},
\end{align}
from where we find
$
\tl{p}_r(x,s) = \fr{\int_{0}^{\infty}p_0(x,u)\,\mathcal{L}\left\{R(t)h(u,t)\right\}du}{1-\tl{r}(s)}.
$
Moreover, the MSD can be obtained from
$
\langle\tl{x}^2_r(s)\rangle
= \fr{\int_{0}^{\infty}\langle x^2(u)\rangle_0\,\mathcal{L}\left\{R(t)h(u,t)\right\}du}{1-\tl{r}(s)},
$
where $\langle x^2(u)\rangle_0=2u$ is the MSD for normal diffusive process.

Let us now consider a subdiffusive process governed by time fractional diffusion equation of order $\alpha$ with $\gamma(t)=\fr{t^{-\alpha}}{\Gamma(1-\alpha)}$, $0<\alpha<1$. The corresponding subordination function is $\tl{h}(u,s)=s^{\alpha-1}e^{-us^{\alpha}}$,
and is inverted to get the time domain representation, that is, $h(u, t) = t g_\alpha(u, t)/(\alpha u)$, where $g_\alpha(u, t) = \mathcal{L}^{-1}\{e^{-u s^\alpha}\} = u^{-1/\alpha} g_\alpha(t u^{-1/\alpha})$ is the one-sided Lévy stable distribution \cite{KGorska21}.
This allows us to write the MSD for subdiffusive process in presence of stochastic resetting
\begin{align}
\label{msd sub reset}
\langle\tl{x}^2_r(s)\rangle =\fr{2}{\Gamma(1+\alpha)}\fr{\mathcal{L}\left\{t^{\alpha}R(t)\right\}}{1-\tl{r}(s)},
\end{align}
where we apply the Mellin transform of the Fox $H$-function, see Section 2.2.3 in \cite{saxena_book}, to evaluate the
integral. Note the role played by the subordination function, wherein the subdiffusive process is now analyzed in terms
of a Gaussian diffusive process, rather than a direct approach like in Refs.~\cite{miquel, nowak}. In a similar way,
the PDF under restarts is
\begin{align}
\label{prks sub reset}
\tl{p}_r(x,s)
=\fr{1}{2|x|}\fr{\mathcal{L}\left\{R(t)H_{1,1}^{1,0}\left[\frac{|x|}{\sqrt{Dt^{\alpha}}}\left|\begin{array}{c c}
    (1,\alpha/2) \\
    (1,1)
  \end{array}\right.\right]\right\}}{1-\tl{r}(s)},
\end{align}
where we apply the Mellin transform of a product of two Fox $H$-functions, see Section 2.3 from \cite{saxena_book}, alongwith
the Legendre duplication formula. It is to be noted that so far we have not made any assumptions about the nature of the resetting
protocol, and (\ref{msd sub reset}) and (\ref{prks sub reset}) represent the general form of MSD and PDF for a subdiffusive
process under stochastic resetting, respectively.

In order to get an understanding of the behavior of a subdiffusive process under stochastic resetting, let us assume $r(t) = r e^{-rt}$, where $r$ is the rate of resetting (all moments of $r(t)$ exist, $\tl{r}(s) = 1 - s/r + s^2/r^2 + \dots$). This implies $R(t) = e^{-rt}$. Using this in (\ref{prks sub reset}) and evaluating the
Laplace transform of the Fox $H$-function \cite{saxena_book} we have
\begin{align}
\label{prxt_Hf}
p_r(x,t) = \int^t_0 dt' ~\fr{e^{-rt'}}{2|x|t'} H_{1,1}^{1,0}\left[\frac{|x|}{\sqrt{Dt'^{\alpha}}}\left|\begin{array}{c c}
    (0,\alpha/2) \\
    (1,1)
  \end{array}\right.\right],
\end{align}
which is the probability of finding a subdiffusing particle at location $x$ at time $t$, resetting to its initial location at
a rate $r$. This allows us to evaluate the steady state density profile $p_{r,st}(x) = \lim_{t \rightarrow \infty} p_r(x,t)$
and equals the Laplace transform $H$-function defining $p_r(x,t)$. Calculating the Laplace transform and simplifying the
resulting form of $H$-function we have
\begin{align}
\label{pr_st}
    p_{r,st}(x) = \fr{\sqrt{r^{\alpha}/D}}{2}\exp\left(-|x|\sqrt{r^\alpha/D}\right),
\end{align}
which has also been derived in earlier works \cite{campos, miquel, nowak,st_we}.

The integral representation of the probability distribution in (\ref{prxt_Hf}) allows us to address the relaxation towards steady state.
In order to achieve this goal, we employ the asymptotic approximation of $H$-function for large arguments since we consider large times and fixed $|x|/t$ \footnote{For large $z$,
$H^{1,0}_{1,1}(z) \sim z^{\beta/\mu}\exp(-\mu C^{1/\mu} z^{1/\mu})$, where $\beta=b_1-a_1+1/2$, $\mu = B_1 - A_1$ and $C = {A_1}^{A_1} {B_1}^{-B_1}$ \cite{ding}.}. From Eq.~(\ref{prxt_Hf}) thus we have 
we have
\begin{align}
\label{asymp}
p_r(x,t) \approx \int^1_0 d\tau ~e^{-t\phi(x,\tau)},
\end{align}
where $\tau = t'/t$, $\phi(r,t') = r\tau + \fr{g(x,\alpha)}{t^{2/(2-\alpha)}}\tau^{-\alpha/(2-\alpha)}$ and $g(x,\alpha) = \fr{2-\alpha}{2}(\fr{\alpha}{2})^{\alpha/(2-\alpha)}(|x|/\sqrt{D})^{2/(2-\alpha)}$. The integral in (\ref{asymp}) can be evaluated using the
Laplace approximation as $p_r(x,t) \sim \exp[-t\phi(r,\tau_0)]$, where the extremum point $\tau_0$ is defined as $\fr{d}{d\tau}\phi(r,\tau)|
_{\tau = \tau_0} = 0$, provided $\tau_0 \in (0,1)$. On the other hand when the extremum point lies outside the unit interval, the maximal contribution
to the integral in (\ref{asymp}) comes from the upper limit of the integral, that is, $\tau = 1$. This allows us to write the large deviation
form for the distribution, that is, $p_r(x,t) \sim \exp[-tI_r(|x|/t)]$ where the rate function is
\begin{align}
\label{ldf_rate}
I_r\left(\fr{|x|}{t}\right) =
\begin{cases}
\sqrt{\fr{r^\alpha}{D}}\fr{|x|}{t}, ~~~\tau_0 < 1,\\
r + \fr{2-\alpha}{2}\left(\fr{\alpha}{2}\right)^{\fr{\alpha}{2-\alpha}}\left(\fr{|x|/t}{\sqrt{D}}\right)^{\fr{2}{2-\alpha}},
 ~~~\tau_0 > 1.
\end{cases}
\end{align}
Comparing (\ref{pr_st}) with (\ref{ldf_rate}) we find that at large times relaxation is achieved in a spatial region demarcated
by $\tau_0 < 1$, while outside this region the subdiffusive walk is relaxing with the front propagating superdiffusively. It is
to be noted that using $\alpha = 1/2$ in (\ref{ldf_rate}) reproduces the rate function governing relaxation
of a random walk on comb under stochastic resetting \cite{rks}. In other words, (\ref{ldf_rate}) governs the relaxation behavior
of a subdiffusive walk under stochastic resetting in arbitrary geometries.

It should be further noted that the Laplace form of NESS for exponential resetting is not limited to power law decaying form of $\gamma(t)$. This is easily seen from Eqs.~(\ref{sub integral}) and (\ref{prks}) in the long time limit as
\begin{align}
\label{prks_general}
p_{r,st}(x)&=\lim_{s\rightarrow0}s\tl{p}_r(x,s)
=\fr{\alpha_r}{2}e^{-\alpha_r|x-x_0|}
\end{align}
where $\alpha_r = \sqrt{r\tl{\gamma}(r)/D}$. It is also evident that the MSD saturates to $\langle x^2_r(t)\rangle\sim 2\alpha^2_r$, in agreement with the recently obtained result in \cite{st_we}.

The above examples show that a subdiffusive process defined by a subordinated Gaussian approaches
Laplace distribution as steady state under exponential resetting. However, in order to address the
existence of a steady state we consider a power-law resetting with
$r(t)=\frac{\gamma r}{(1+r t)^{1+\gamma}}$, with $\gamma>0$ \cite{mendez_power}, for which $\hat{r}(s)=\gamma U(1,1-\gamma ,s/r)$, where $U(a,b,z)$ is the Tricomi confluent hypergeometric function, and $R(t)=\frac{1}{(1+r t)^{\gamma}}$. Thus, the MSD becomes
\begin{align}
\label{msd sub pareto reset}
\langle\tl{x}^2_r(s)\rangle 
=\fr{2}{r^{1+\alpha}}\fr{U\left(1+\alpha,2-\gamma+\alpha,s/r\right)}{1-\gamma U(1,1-\gamma ,s/r)}.
\end{align}
For $t\rightarrow \infty$ we have the behavior
\begin{align}
    \langle x^2_r(t)\rangle\sim\left\{\begin{array}{ll}
        \fr{2\Gamma(\alpha+1-\gamma)}{\Gamma(1-\gamma)\Gamma^2(1+\alpha)}t^{\alpha}, \quad & 0<\gamma<1,  \\
        \fr{2}{\alpha\Gamma(1+\alpha)}t^{\alpha}/\ln{t}, \quad & \gamma=1,  \\
        \fr{2(\gamma-1)}{r^{\gamma-1}(\alpha+1-\gamma)\Gamma(1+\alpha)}t^{\alpha+1-\gamma}, \quad & 1<\gamma<1+\alpha, \\
        \fr{2\alpha}{r^\alpha \Gamma(1+\alpha)}\ln{t}, \quad & \gamma = 1+\alpha, \\
        \fr{2}{r^\alpha}\fr{(\gamma-1)\Gamma(\gamma-1-\alpha)}{\Gamma(\gamma)}, \quad & \gamma>1+\alpha.
    \end{array}\right.
\end{align}
This implies that at long times MSD for subdiffusion under power-law resetting saturates to a constant value only when $\gamma>1+\alpha$. This is in sharp contrast to diffusion with resetting for which the MSD saturates only if $\gamma>2$, i.e., both the mean and variance exist. This surprising result can be understood as follows. In presence of resetting
the system is composed of the random walk and the resetting protocol, and it
will relax to a steady state only when fluctuations become constant in time. Now this can be achieved either by reducing fluctuations in the system (measured by MSD) at the cost of increased fluctuations in reset times (non-existence of moments) or vice-versa. The cases of diffusion and subdiffusion discussed above corroborate this assertion. In other words, diffusion involves linear dependence of MSD and hence requires existence of both finite
mean and variance in the distribution of reset times in order to achieve a steady state. On the other hand, for
subdiffusion MSD behaves as $t^{\alpha}$ with $0<\alpha<1$ and thus the NESS can be achieved by only by requiring
the existence of a finite mean. In this respect, for a process for which MSD saturates to a constant value without resetting, the behavior of the distribution of reset times becomes immaterial since the system will anyway relax to the steady state no matter if the distribution of reset times possesses finite mean and variance or not. It should also be noted that for $\alpha=1$ we recover the results for diffusion under resetting. 

\textit{Discussion}:
The ubiquitous occurrence of diffusive transport in statistical physics implies that a general theory addressing the effect
of stochastic resetting on diffusion is need of the hour. We study the long term behavior of diffusion under stochastic
resetting and find that MSD saturates to a constant value only when the distribution of reset times possesses both mean and
variance. For the case when variance does not exist, diffusion turns into subdiffusion and the extreme case when mean does
not exist the transport remains diffusive, though with a reduced value of diffusive coefficient. For all these cases
we derive the leading order contribution to the PDF of a Gaussian propagator under restarts, which exhibits a cusp
at the location of reset when the distribution of reset times possesses both a finite mean and variance.

Diffusive transport is, however, a special case of 
anomalous diffusion which rules transport in heterogeneous media. Motivated by this observation we study the effect of stochastic
resetting on subdiffusion and derive exact expression for probability distribution using subordination in Laplace space for
an arbitrary resetting protocol. For the case when resetting takes place at a constant rate, we are able to invert the Laplace
transform and find the PDF in terms of $H$-function. Similar to the case of a Gaussian diffusive propagator we find that
the steady state of subdiffusive motion is also a Laplace distribution. We also derive the large deviation form of subdiffusive
PDF under resetting at constant rate. We find that the steady state builds from the location of reset with the relaxation
front propagating outwards in a superdiffusive manner. This generalizes the previous result addressing stochastic resetting
on combs to arbitrary value of subdiffusion exponent $\alpha$.
We also show that the existence of a NESS for subdiffusion requires the distribution of reset times to possess only
a finite mean, in sharp contrast with diffusive process which requires existence of both finite mean and variance.
In other words, a stochastic process under stochastic resetting can achieve a NESS by balancing the intrinsic
fluctuations against fluctuations in reset times and vice-versa.

\textit{Acknowledgments}:
RKS thanks IASH and CHE Excellence Fellowship for postdoctoral research. RKS thanks Shlomi Reuveni for critical comments and insightful discussions. RKS thanks Shamik Gupta for useful discussion regarding power-law restarts. RKS thanks Dibyendu
Das for useful comments. KG's research was supported by the NCN Research Grant OPUS-12 No. UMO-2016/23/B/ST3/01714 as well as by the NCN-NAWA Research Grant Preludium Bis 2 No. UMO-2020/39/O/ST2/01563. TS was supported by the Alexander von Humboldt Foundation. TS acknowledges financial support by the German Science Foundation (DFG, Grant number ME 1535/12-1).

\end{document}